\documentclass[journal]{IEEEtran}
\usepackage{graphicx}
\usepackage{tikz}
\usepackage[noadjust]{cite}
\usepackage{tikzpagenodes}
\usepackage{hyperref}



\begin{document}
\title{Silicon Photonics in Optical Access Networks for 5G Communications}

\author{Xun~Guan,~\IEEEmembership{Member,~IEEE}, Wei~Shi,~\IEEEmembership{Senior Member,~IEEE}, Jia~Liu, Peng~Tan, Jim~Slevinsky,~\IEEEmembership{Senior Member,~IEEE}, and
        Leslie~A.~Rusch,~\IEEEmembership{Fellow,~IEEE}
        }

\maketitle
\begin{tikzpicture}[remember picture,overlay]

\node[align=right,text=blue] at ([xshift=10em,yshift=2em]current page text area.north) {\href{https://youtu.be/AfT_0g3hPhA}{\LARGE{\underline{COMPANION YOUTUBE VIDEO}}}};

\node[align=right,text=blue] at ([xshift=-16.5em,yshift=-2em]current page text area.south) {\href{https://youtu.be/AfT_0g3hPhA}{YouTube link: \nolinkurl{https://youtu.be/AfT_0g3hPhA}}};
\end{tikzpicture}%

\begin{abstract}
Only radio access networks can provide  connectivity across multiple antenna sites to achieve the great leap forward in capacity targeted by 5G. Optical fronthaul remains a sticking point in that connectivity, and we make the case for analog radio over fiber signals and an optical access network smart edge to achieve the potential of radio access networks. The edge of the network would house the intelligence that coordinates wireless transmissions to minimize interference and maximize throughput. As silicon photonics provides a hardware platform well adapted to support optical fronthaul, it is poised to drive smart edge adoption. We draw out the issues in adopting our solution, propose a strategy for network densification, and cite recent demonstrations to support our approach. 
\end{abstract}


\IEEEpeerreviewmaketitle

\section{Introduction}
\IEEEPARstart{A}{s} digital cellular networks have evolved into the fifth-generation (5G), industry is predicting that worldwide 5G deployment will be faster than preceding generations due to growing demand for ubiquitous connectivity and high bandwidth. We examine the strategic combination of three technologies (a smart edge, analog radio over fiber, and micro-resonators in silicon photonics) to meet 5G latency and densification goals. 

The smart edge is located close to radio towers to meet latency requirements in 5G coordination and in applications such as autonomous driving, cellular vehicle-to-everything (C-V2X) applications, cloud gaming, analytics for big data and artificial intelligence, as well as numerous cloud-native mobile applications. The latency must be combined with high bandwidth as 5G network should support media and entertainment verticals in an efficient and scalable way \cite{TELUS2}. Current 360-degree video 4K needs about 20-50~Mbps for a satisfactory streami ng experience, and 8K needs at least 100~Mbps. Strong-interactive virtual reality real-time responses to user actions impose a latency below 10~ms; stereo vision require 200+~Mbps throughput. To reach the human vision limit, requirements could climb to 5.2~Gbps (theoretical calculation with 30K$\times$24K resolution and 120 frames per second).

\begin{figure}[!b]
\begin{center}
    \includegraphics[width=.48\textwidth]{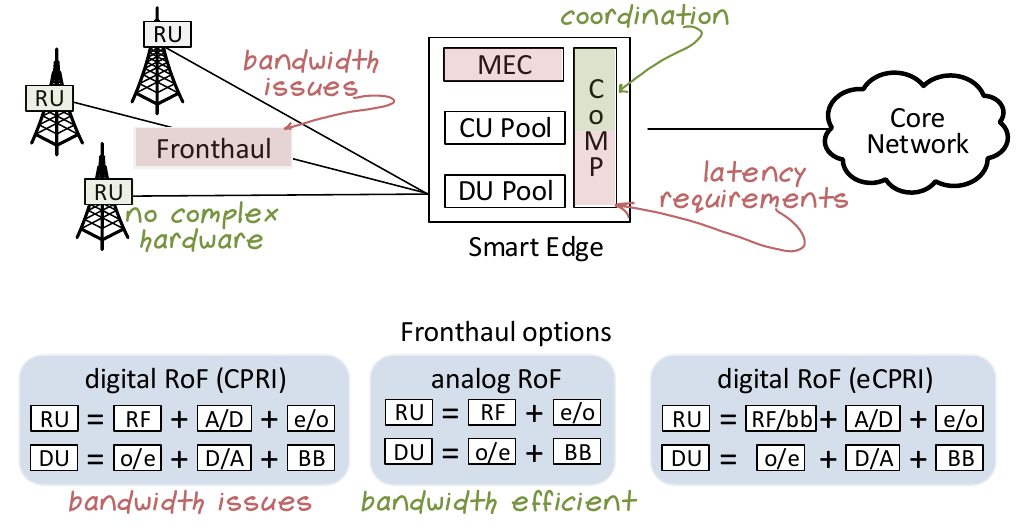}
 \caption{Proposed RAN architecture for optical fronthaul; A/D analog-to-digital, D/A digital-to-analog, o/e opto-electronic, e/o electro-optic, RF radio frequency, BB or bb baseband.. }
  \label{archtectures}
\end{center}
\end{figure}

To simultaneously meet bandwidth and latency requirements, we need spectrally efficient fronthaul that carries physical (PHY) layer RF signals to the smart edge for coordinated signal processing. This radio access network (RAN) linking tower site(s) to the network edge should be dense, and therefore radio head hardware should be simple to keep cost down. Analog radio over fiber meets these requirements. Densification can be enhanced with support of wavelength division multiplexing (WDM). 

We have conjectured a novel migration of deployed fiber infrastructures to support WDM for heterogeneous services. In our early stage research, we have examined an integrated solution in silicon photonics with microring resonators. The resonator diminutive size can support many wavelength channels in very small chip real estate, with each channel consuming low power per bit \cite{dube2016ultrafast}.  Our demonstrations lay the foundation for pushing these concepts even farther as we move from 5G to 6G and we hope that this will motivate further work along this direction to make our vision a reality.

\section{How do we evolve to a RAN Smart Edge?}

RAN architectures for 5G must support elastic interfaces and flexible deployment, including the option of multiple functional splits  \cite{TELUS3}. In 4G all baseband signal processing was concentrated in a base band unit (BBU), while 5G can split the same functionality across a central unit (CU), distributed unit (DU) and remote unit (RU). This strategy works in concert with multi-access edge computing (MEC) that puts data-center type capabilities at the edge of the network for latency-sensitive applications. Figure~\ref{archtectures} illustrates our proposed scenario for RAN deployment. We will first describe alternative scenarios and their weaknesses to highlight the advantages of our proposal.

The traditional solution has all equipment (RU, MEC, CU and DU) at the remote antenna site.  There are no latency issues (all processing is on-site) and no fronthaul (no transport to a remote processing site). However, it comes with two large disadvantages: expensive hardware at locations difficult to service and no 5G coordination across cells.  It scales very poorly for dense cell sites.

Another functional split has all RF baseband processing is at the remote site (RU and DU) and a smart edge with the MEC and a CU. Proximity of the smart edge allows the MEC to support applications with strict latency requirements; and the central unit (CU) supporting upper layer data also enjoys low latency. The mid-haul from antenna sites to the smart edge carries only upper layer data (no wireless PHY signals) and only imposes moderate bandwidth requirements. As in the first scenario, there can be no 5G RF coordination. 

Our proposal is a possible embodiment of option~8 in \cite{TELUS3}: wireless PHY functionality located in a distributed unit pool at the smart edge (co-located with the MEC and CU pool). The separation of baseband processing (at the DU) and RF chains (at the RU) enables the move to an architecture where commodity hardware (at locations with easy accessibility and maintenance) can be used for baseband processing. The box marked CoMP, explained in the next section, is one of the keys to increased 5G performance. The challenges for this architecture lie in managing fronthaul bandwidth.

\section{What can we gain with a Smart Edge?}
\label{trends}
Within a long-reach access network, a feeder fiber of 100~km has inherent 1~ms latency, fatal to latency-sensitive 5G services.
The 200~$\mu$s round-trip latency of a 20~km tributary fiber is a reasonable choice for locating a processing site, i.e., the ``edge'' of the network. 

\textbf{Edge computing:}  Key applications in 5G, including virtual reality, augmented reality, tactile networks and C-V2X, in many cases must have latency be as low as 1~ms, and reliability in excess of 99.999\%. By co-locating the central unit with the MEC at the smart edge, network layer processing can also be streamlined. 
The IMT-2020 standard specifies 5G should meet 100~Mbps downlink and 50~Mbps uplink in dense urban areas, and 20~Gbps downlink and 10~Gbps uplink peak data rate in enhanced mobile broadband (eMBB) \cite{TELUS1}. The minimum requirement for one way user plane latency is 4~ms for eMBB and 0.5~ms for ultra-reliable low latency communications (URLLC).

\textbf{Network Densification:}
Both spatial densification and spectral aggregation are necessary for 5G \cite{bhushan2014network}. Spatial densification involves the deployment of ultra dense picocells for ubiquitous coverage, while spectral aggregation complements legacy spectra with new radio capacity.
The number of access points (antenna sites) must increase significantly for densification, making expensive hardware solutions for remote units inappropriate. As RF signals climb higher in the spectrum and as wireless signals become wider band, great pressure is applied to bandwidth requirements in the fronthaul network. 

Consider fronthaul solutions at the bottom of Fig.~\ref{archtectures} and their impact on densification. Two options currently exist: a broadband digital optical connection employing the conventional common public radio interface (CPRI), and an updated version known as enhanced CPRI (eCPRI). The excessive CPRI bandwidth expansion \cite{TELUS4} (an order of magnitude above occupied RF bandwidth)  led to eCPRI. Despite improved spectral efficiency, eCPRI incurs significant processing and queue delay, as well as a drastic increase of RU complexity (note the ``bb'' in Fig.~\ref{archtectures}  indicating some baseband processing it located here) and cost \cite{giannoulis2018analog}. For these reasons, in this paper we focus on the dedicated fiber wireless (FiWi) tunnel, i.e., analog radio-over-fiber (RoF) as opposed to the CPRI and eCPRI digital RoF \cite{lim2009fiber} solution.

\textbf{RF Coordination:}
Densification to increase capacity also implies smaller cells where signal overlap among cells is unavoidable. To improve system throughput the inter-cell interference should be combated or harnessed by cooperative multipoint (CoMP) technologies \cite{zhang2017experimental}. The coordination between antenna sites is key, either by centralized or distributed controllers. Coordination requires inter-connectivity via an optical access network (OAN) as only optics can provide the necessary bandwidth for fronthaul. 

Joint processing algorithms among cooperating sites usually require  synchronization within $\pm$1.5~$\mu$s \cite{TELUS6}, and latency less than 150~$\mu$s to ensure channel state information can be exploited. The RAN of scenario~3 in Fig.~\ref{archtectures} would be sufficiently proximate and house sufficient hardware resources for CoMP to exploit signals from several RUs within these tolerances.
Locating the CoMP controller closer to the remote antenna sites  reduces fundamental latency, and  equipment is located on network operator premises (remote node) which are highly reliable, controllable, and accessible. %

\section{How can we exploit a deployed PON for RAN?}

Overlaying 5G signals on  existing  networks allows increased revenue from an already installed fiber network, as well as assisting in network densification. Subsystem integration is clearly a path to reducing upgrade hardware cost in 5G OANs. In this paper we focus on silicon photonics (SiP) solutions that are integrable, and  resonant structures that are small-footprint and low-power.

\subsection{Proposed Architecture}
Access networks, such as passive optical networks (PON) and fiber-to-the-x (FTTX), typically use a tree hierarchy. A main feeder line runs from the central office to neighborhoods where fiber branches out. We  adapt this architecture to accommodate 5G fiber links in the distribution branches. We allow for compatibility with WDM, as it will likely be used to boost the overall throughput in future deployments.

\begin{figure}
  \includegraphics[width=0.48\textwidth]{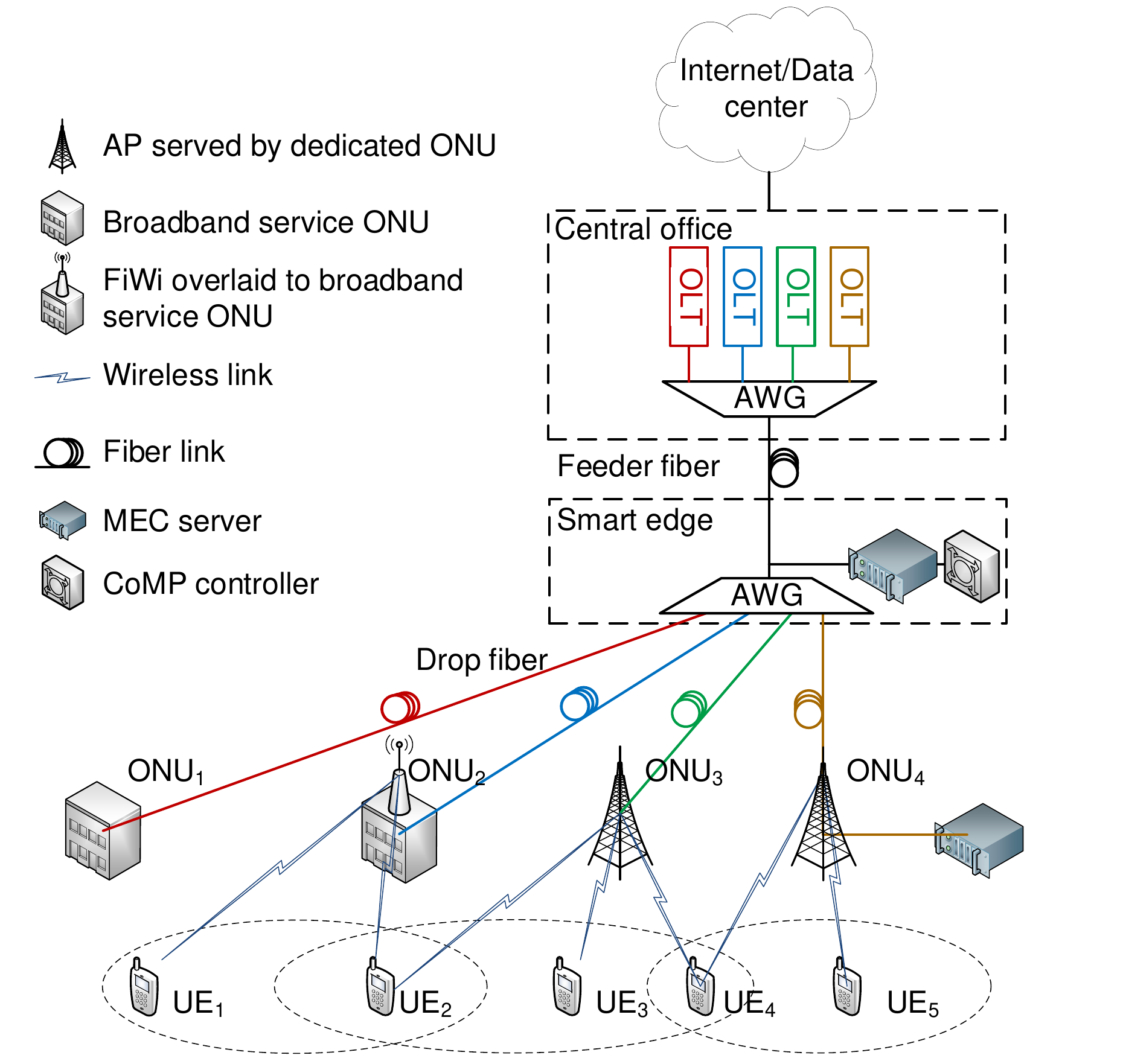}
  \caption{A typical optical access network. For illustrative purpose, four ONU represent four different scenarios from left to right: (1) broadband service only; (2) FiWi overlay on broadband service; (3) dedicated wireless access point; and (4) access point with an MEC server}
  \label{PON}
\end{figure}

In Fig.~\ref{PON} we propose an OAN  preserving the PON tree-topology with minor hardware upgrades to accommodate 5G signals. For the sake of discussion and illustration, four WDM bands are assumed. We consider different use cases for each WDM channel.  Intelligence at the remote node provides cloud-like services (via the MEC server), and enhanced 5G processing (via the CoMP control). 
In the next section we describe the SiP solutions (at the remote node and at the ONUs)  allowing 5G overlay without  significant degradation of the original broadband services. 

In Fig.~\ref{PON} four ONU use cases are illustrated. The first, ONU\textsubscript{1}, is a conventional link providing broadband service in FTTX. 
ONU\textsubscript{2} offers broadband service like ONU\textsubscript{1}, but with a remote unit (RU) embedded to provide 5G services to users in the vicinity. 
ONU\textsubscript{3} is a dedicated fronthaul for a 5G RU, with no digital services. As with ONU\textsubscript{2}, ONU\textsubscript{3} fronthaul signals are delivered to the smart edge. In ONU\textsubscript{4}, a dedicated MEC server deployed at the RU leaves the optical fronthaul dedicated to CoMP signals. Wireless user equipment (UE) is scattered throughout the overlapping coverage areas of the RUs. We note that backhaul (for 5G signals) is a digital PON service on the feeder fiber.

The distribution of RUs and UEs in Fig.~\ref{PON} illustrates how the CoMP unit at the smart edge can exploit the RAN. The RUs at the ONU\textsubscript{2} and ONU\textsubscript{3} can execute CoMP beamforming, waveform precoding, or MIMO strategy, thus avoiding or harnessing interference among the UEs. This cooperation relies on the availability of raw 5G signals from ONU\textsubscript{2} and ONU\textsubscript{3} routed to the smart edge. 

\subsection{Analog overlay advantages}
An overlay maximizes the benefit of reusing the deployed infrastructures. The RU can be mounted at conventional FTTX facilities as in ONU\textsubscript{2}, or at dedicated antenna sites as in ONU\textsubscript{3}. As described in the fronthaul options in Fig.~\ref{archtectures}, the complexity of the RU is much lower for our proposed analog RoF overlay. In our proposal RU hardware is simply an RF chain and electro-optic (downlink) or opto-electric (uplink) conversion. Power consumption is reduced without analog-to-digital (ADC) and digital-to-analog (DAC) conversion in downlink and uplink, respectively.

In eCPRI, part of the baseband signal processing function is relocated to the RU, to alleviate the bandwidth limitation. However, this  greatly increases the RU cost. In contrast, analog RoF needs no complex RU processing units, significantly reducing the cost of massive deployment. It also avoids the bandwidth limitation in CPRI and eCPRI, and shows high scalability when more bandwidth is needed.

\section{Can silicon photonics  enable the Smart Edge?}

\begin{figure*}
\begin{center}
      \includegraphics[width=.9\textwidth]{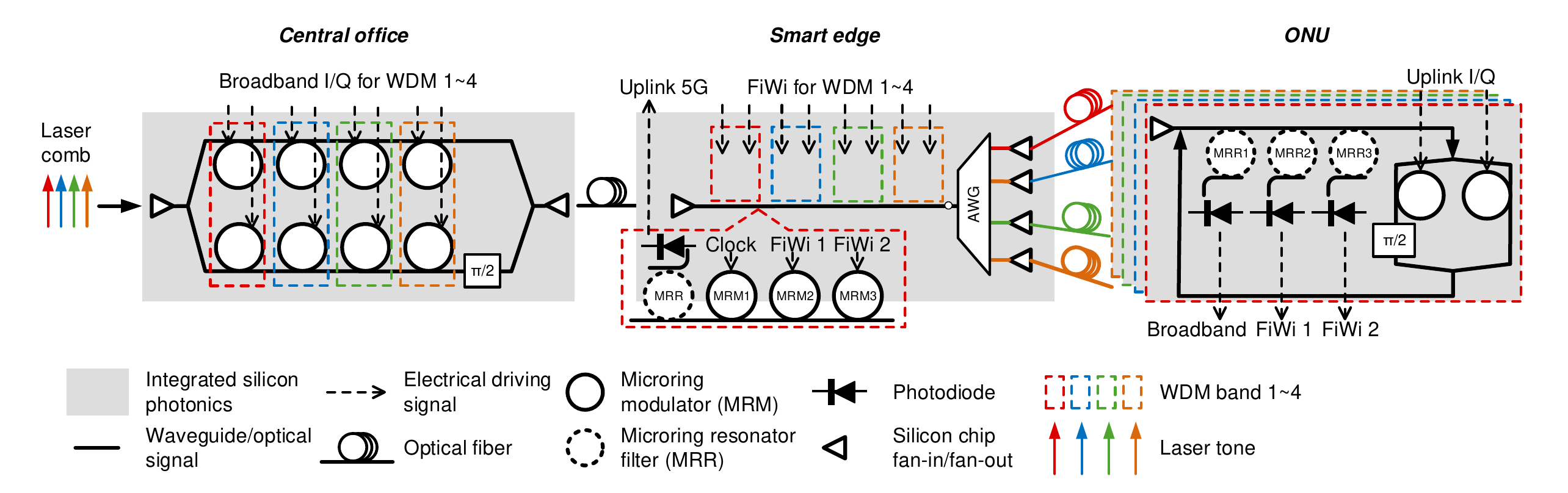}
  \caption{The proposed framework.}
  \label{solution}
\end{center}
\end{figure*}
Optical subsystems enable the coexistence of broadband digital signals and 5G signals in the distribution network, but  require new, unexplored hardware at the smart edge and the ONU. 
We propose the use of a microring resonate structures on a bus. A microring resonator (MRR) filters the passband \cite{bogaerts2012silicon}, while a modulator (MRM) is tuned to modulate one wavelength and to let all other wavelengths pass unmodified. MRMs are among the smallest modulators available, enabling a small footprint and low power consumption even as we scale to more and more WDM channels \cite{dube2016ultrafast}. 
In Fig.~\ref{solution} we provide block diagrams of subsystems  for the proposed architecture. The grey shading indicates SiP blocks. 

\subsection{Central Office}
The central office optical line terminal (OLT) equipment has multiple laser tones for WDM modulated by the SiP subsystem. 
Each MRM illustrated in Fig.~\ref{solution} at the central office is identical, and tuned to a particular wavelength (tone). The SiP chip has two branches (I and Q) to accommodate complex modulation; a phase shifter generates a $\pi / 2$ phase difference between branches.  In our illustration, a four-wavelength comb passes through four MRM pairs; each pair modulates one tone. The working wavelength is selected by thermal tuning. The IQ modulators can create a single sideband real signal to enable inexpensive direct detection at the ONU and to avoid dispersion-induced fading.

Unlike MRMs, non-frequency-selective silicon Mach-Zehnder modulator would require arrayed waveguide gratings (incurring significant power loss) to separate tones before modulation and to combine them after modulation. The MZM waveguide length is an order of magnitude longer than the MRM, increasing the footprint and leading to power consumption of picojoules in an MZM versus femtojoules \cite{dube2016ultrafast} in an MRM. 

\subsection{Smart edge}
In Fig.~\ref{solution} the site of branching from the feeder fiber to the end users would be called a remote node in legacy systems (an AWG for WDM PONs); further power splitting within wavelengths could also be present. The smart edge moves away from the totally passive paradigm, hence our referring to the proposed architecture as an OAN instead of a PON, and the labeling “smart edge” for the branching point.

Downlink digital signals to end users are unchanged, with 5G signals added in unoccupied sections of the WDM slot. In the uplink, digital signals are unchanged and 5G signals dropped at the smart edge. With 50~GHz WDM slots, for example, we could reserve 20~GHz for digital data and the remainder for 5G signals. The 20~GHz subband for high spectral efficiency, high bit rate signaling could be time multiplexed when passive splitting is used in the network. We could frequency multiplex 5G signals for multiple antenna RUs. 

As illustrated in Fig.~\ref{solution}, our SiP solution introduces cascaded MRMs (bus configuration) between the fiber feeder and the AWG.  Recall that wavelength selective MRMs in a bus allow us to tune each MRM to work at a specific wavelength, leaving other WDM bands untouched. At each WDM channel, we assign three MRMs to generate a 5G downlink and one MRR to intercept a 5G uplink. Four our four-channel downlink example in Fig. \ref{PON}, we would cascade a series of 12 MRMs. The SiP benefits for the central office apply to the smart edge: colorless, broadband, etc. 

Out of the three MRMs in each group, the first MRM generates two subcarriers by modulation of one laser tone by a clock signal at $f_{s}$.  Subcarrier separation can be easily adjusted via the clock frequency. The second and third MRMs are tuned to the wavelengths of the two newly generated subcarriers. Two independent 5G service tunnels are constructed in each WDM band, opening new spectrum without high bandwidth modulators and paving the way for new 5G services.

\subsection{ONU}
We next turn to the customer premises equipment in Fig.~\ref{solution}. Our optical network unit (ONU) must now receive  5G signals in addition to the downlink broadband digital data. The ONU receiver has cascaded MRRs to strip off the broadband signal and one or two 5G service tunnels with analog RoF signals. A single IQ MRM modulator creates the uplink signals; the digital uplink and RoF signal could be mixed before electro-optic conversion. Alternatively, a three MRM bus (as in the smart edge could be used to avoid RF mixing. 

The ONU is totally colorless and adapts to the wavelength delivered in the downlink. As there is no local laser source, the downlink carrier is used both as the beat tone for direct detection and as the carrier for uplink generation. Higher beat tone power leads to better performance, but the total available power is limited and must be carefully husbanded. Sufficient power is apportioned to photodetectors to assure reliable reception of all signal types (digital and RoF). 
The residual carrier for the uplink (broadband and 5G) transmission is diminished, however, the central office (where laser sources are easily available) can exploit coherent detection to achieve comparable link performance. SiP coherent receivers would make this cost effective.

As seen in Fig.~\ref{solution}, at the first stage of the SiP ONU system we use three identical higher order microring resonator (MRR) filters to strip off each signal sub-channel. We tune the passband center wavelength of the colorless MRRs  by thermal control (as in the smart edge MRMs). Using higher order MRRs \cite{chen2014high} allows us to thermally tune the bandwidth as well as the center wavelength. We drop the broadband signal with MRR1, tuning the passband to encompass the signal and only an appropriate portion of carrier power for the beat tone for detection. Each of the following MRRs is similarly tuned to achieve good 5G signal integrity. The three dropped signals are each direct detected with on-chip photodiodes compatible with silicon technology.

Dropped downlink signals are suppressed on the bus. The residual carrier  passes to an IQ MRM similar to those in the OLT. We can place the uplink SSB signal spectrum on the opposite side of the central carrier. Avoiding overlapping of downlink and uplink spectra decreases  Rayleigh backscattering. The uplink signal is then fed back to the OAN fiber.

\section{Where do we stand now?}

\subsection{Adding RoF to downlink traffic}
In \cite{guan2021heterogeneous} we presented experimental results where we intercept the downlink digital wideband signals at the smart edge and add  5G signals. The overlaid analog RoF combined with the digital signal is sent to the local distribution network.
We used a series of cascaded modulators that we designed, had fabricated. 
In this demonstration, two WDM bands at 100~GHz separation are included. The digital signals were 10~Gb/s. The feeder fiber before the smart edge was 20~km, and the distribution fiber was 5~km.
\begin{figure}[bht]
\begin{center}
  \includegraphics[width=0.4\textwidth]{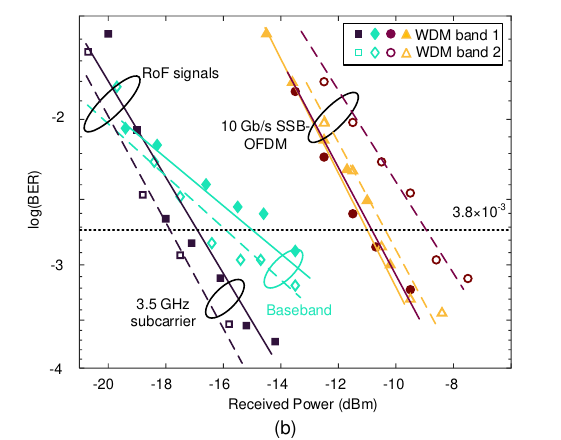}
  \end{center}
  \caption{Measured smart edge 
  bit error rate for various signals from  \cite{guan2021heterogeneous}. }
  \label{spec}
\end{figure}

MRM1 generates two subcarriers in the first WDM band from a 20~GHz clock signal. MRM2 modulates one subcarrier with baseband OFDM   that could be upshifted to the 5G carrier frequency at the RU. Thermal tuning moves  MRM3 and MRM4 to the second WDM band for similar operation. The 5G signal this time was placed on the 3~GHz wireless carrier and could be sent directly to the antenna after opto-electronic conversion at the RU. A real-time oscilloscope and offline processing emulated the ONU functions. We validated good performance of all signals as reported in \cite{guan2021heterogeneous}, with the bit error rate  (BER) measurements given in Fig.~\ref{spec}(b).

\subsection{Detecting RoF and remodulating the uplink}
In \cite{Lyu_OE_2020} we reported a demonstration of essential functions in the ONU.
Our new receivers for the customer premises separate and independently detect the digital Internet service signals and the 5G RoF signals. 
We designed and had fabricated the SiP chip for  the ONU subsystem described in Fig.~\ref{solution}.

We used a signal generator to create the smart edge combination of digital broadband and analog RoF signals on one WDM channel.  The analog RoF detection included five 125~MHz RF signals spaced at 250~MHz, and the 
broadband detection a 16~Gb/s OFDM (orthogonal frequency division multiplexed) signal. Both analog RoF and broadband digital signals were successfully detected with bit error rate (BER) below the typical 7\% forward error correction (FEC) threshold. The analog RoF reduced the received carrier by 4~dB, while the broadband detection used an additional 2~dB of power margin. The residual carrier was remodulated and successfully detected. In Fig. 5 we present the optical spectrum exiting the ONU. The uplink signal was 13~dB obove any residual downlink signals.

\begin{figure}[bt]
\begin{center}

  \includegraphics[width=0.42\textwidth]{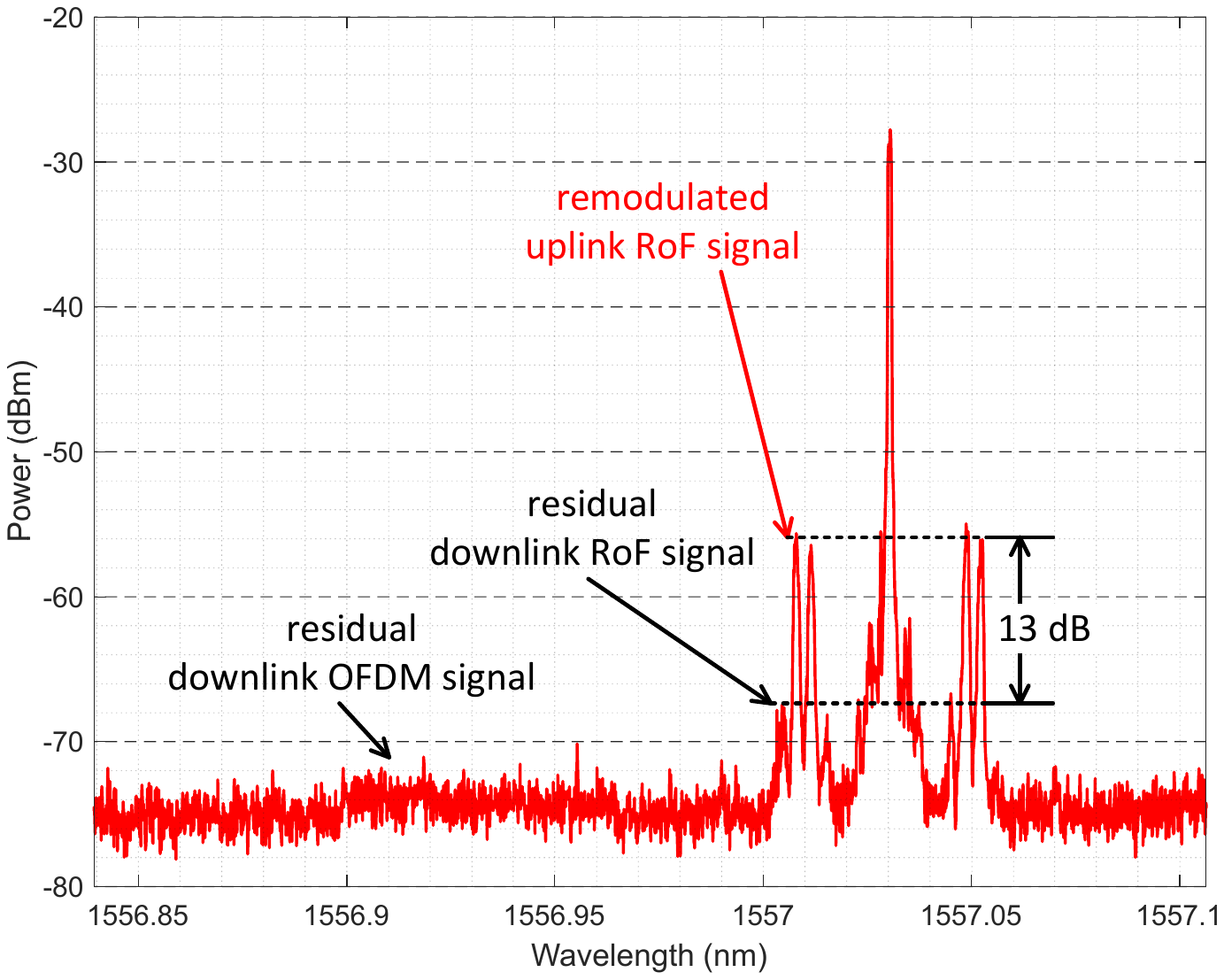}
  \caption{Optical ONU spectra showing remodulated uplink  and suppressed downlink signals \cite{Lyu_OE_2020}. }
  \label{ONU_chip}

\end{center}
\end{figure}

\section{Is it really that simple?}
We focused on the major issues in SiP subsystems for RANs. We glossed over many technical refinements, using simpler SiP structures than those we would use in practical subsystems. In the following we point out how SiP can be designed to meet those practical challenges.

\subsection{Power Budget and Amplification}
While a PON does not permit any amplification after leaving the central office, this constraint would not apply in an OAN with a smart edge. Our solution is based on an upgrade of the remote node from a repository for splitters and AWGs to a cabinet for MECs and CoMP controllers. In such a scenario, the increase in cost from adding amplification at the smart edge would be more than offset by the revenue from the newly supported 5G services.

Our proof-of-concept demonstrations used bare SiP chips with high coupling loss, requiring additional amplification \cite{guan2021heterogeneous,Lyu_OE_2020}. Commercial packaging would reduce this to a few dB, and  the microrings in the bus introduce low filtering and propagation loss in their passband. Our SiP MRM bus offers good scalability for WDM; the short waveguide length makes incremental loss in adding channels small (loss is dominated by coupling even with efficient commercial packaging); please see \cite{guan2021heterogeneous,guan2021overlaying} for a fuller discussion.

\subsection{Polarization Diversity}
One immediate question when applying silicon photonics to the OAN is the handling of polarization; received signals can have arbitrary polarization, yet SiP is very sensitive to polarization. The solution is to use polarization diversity in our subsystems: polarization insensitive edge couplers followed by a polarization splitter and rotator. The cost of this solution is doubling the number of some passive components. The additional complexity can be minimized by clever design and routing. Overall, the small size of the components make this a tolerable compromise.  

\subsection{Thermal Stabilization}
Microring-based devices are highly temperature-dependent. This characteristics allows us to select the wavelength band via thermal tuning. On the other hand, it necessitates the thermal control of microrings, as even a small temperature change would lead to a drift of response, and a loss of resonance. We are working on simple low-frequency feedback loops to provide highly reliable ring resonator stabilization. We can use the microring structures drop ports to easily channel a monitoring signal to on-chip photodiodes. As photodiodes and electronic feedback circuits can be integrated on the same SiP chip beside the ring resonators, the additional cost and complexity will be manageable.


\section{Conclusion}
We have proposed an architecture to exploit analog radio over fiber in an optical access network. 
A smart edge attacks latency, the analog RoF provides spectrally efficient fronthaul and simple remote unit hardware, while we propose silicon photonics to provide small footprint integration that can scale well as future systems adopt multiple wavelength channels. 

We examined network configurations that allowed the merging of 5G fronthaul and broadband WDM signals in an optical access network.  
The architecture aligns well with principles in classic PONs (tree hierarchy, distributed carrier, remote nodes, etc.), but with an upgrade to a smart edge to achieve the potential of 5G efficient, orchestrated operation. We proposed specific SiP subsystems compatible with an efficient exploitation of WDM spectral resources. 

With the rapid evolution of C-V2X technology, especially exploiting 5G next generation radio, a significant opportunity exists for a logical communications overlay enabling connectivity of road side units enabling an efficient path to provide core connectivity to the peer-to-peer connected vehicle infrastructure.  The proposed overlay would enable a mobile network operator synchronization capability for the C-V2X infrastructure.  Beyond this, virtualization of the road side units themselves is possible for more efficient and resilient utilization of computing and storage resources --- supporting vehicle safety and enabling entertainment applications.

\ifCLASSOPTIONcaptionsoff
  \newpage
\fi






\vskip 0pt plus -1fil

\begin{IEEEbiographynophoto}{Xun Guan}
received the B.E. degree from Huazhong University of Science and Technology in 2012, and the Ph.D. degree from The Chinese University of Hong Kong in 2016. Since 2017, he has worked as Postdoctoral Research Fellow in Centre for Optics, Photonics and Lasers, Université Laval, QC, Canada, where he holds a position of Research Professional now. His research interests include photonics, digital signal processing, silicon photonics and the applications in optical communications.
\end{IEEEbiographynophoto}
\vskip -0.5\baselineskip plus -1fil
\begin{IEEEbiographynophoto}{Wei Shi}
is an Associate Professor with the ECE department, Université Laval, Québec, Canada and holds a Canada Research Chair in Silicon Photonics. He received a Ph.D.  in ECE from the University of British Columbia, Vancouver, Canada.  He  held a Postdoctoral Fellowship from the Natural Sciences and Engineering Research Council of Canada (NSERC)  at McGill University, Montreal, QC, Canada. 
\end{IEEEbiographynophoto}
\vskip -0.5\baselineskip plus -1fil
\begin{IEEEbiographynophoto}{Jia Liu}
received the Ph.D. degree in electrical engineering from the University of Alberta in 2007, Edmonton, AB, Canada. She is with the TELUS Communications Inc. where she defines technology architecture and roadmap of 5G RAN. Jia also leads the virtualized RAN and open RAN evolution. Her research interests are in radio access technologies, edge computing as well as RAN evolution for next generation wireless communication systems.   
\end{IEEEbiographynophoto}
\vskip -0.5\baselineskip plus -1fil
\begin{IEEEbiographynophoto}{Peng Tan}
is with TELUS Communications Inc. where he is currently working on technology strategy and architecture development for 5G and immersive media experience. His current research interests include wireless and wireline convergence support for the 5G system and 5G media streaming architecture.
Peng received the Ph.D. degree in electrical engineering from the University of Alberta in 2006. 
\end{IEEEbiographynophoto}
\vskip -0.5\baselineskip plus -1fil
\begin{IEEEbiographynophoto}{Jim Slevinsky}
received a B.Sc. (EE), and an M.Sc. (ECE), from the University of Alberta in 1983 and 1999, respectively. Jim is Director - Technology Strategy in the Chief Technology Office of TELUS Communications, in Edmonton, Alberta, where he is responsible for discovery, evaluation, and initial implementation of disruptive service and technology suites. Previously, Jim was a Research Scientist at TRLabs, where he developed RingBuilder\textsuperscript{{TM}} an optical network automated synthesis tool. Prior to that, as a member of scientific staff at Bell-Northern Research, Jim developed optical transport and advanced switching systems. Jim is a Senior Member of IEEE, and a member of APEGA.
\end{IEEEbiographynophoto}
\vskip -0.5\baselineskip plus -1fil
\begin{IEEEbiographynophoto}{Leslie A. Rusch} (F)
received the B.S.E.E. degree (with honors) from the Caltech in 1980 and the M.A./Ph.D. degrees in EE from Princeton University in 1992/1994. She holds a Canada Research Chair in Communications Systems Enabling the Cloud and is a full professor in the ECE department at Université Laval, QC, Canada, and is a member of the Centre for Optics, Photonics and Lasers. She received the IEEE Canada Fessenden Award for Telecommunications Research and the IEEE Canada Award for Graduate Supervision and is a Fellow of OSA and IEEE.
\end{IEEEbiographynophoto}
\end{document}